\documentclass[aps,prx,reprint,floatfix,superscriptaddress,footinbib,twocolumn]{revtex4}
\usepackage{xcolor}
\usepackage{microtype}
\usepackage[bookmarks=true,
bookmarksnumbered=false, 
bookmarksopen=false, 
colorlinks=true,
linkcolor=blue]{hyperref}
\definecolor{webblue}{rgb}{0, 0, 0.5} 
\hypersetup{colorlinks=true,urlcolor=blue,citecolor=blue}
\usepackage{graphicx}
\usepackage{subfigure}
\usepackage{url}
\usepackage{hyperref}
\usepackage{inconsolata}
\usepackage{amsmath}
\usepackage[capitalize]{cleveref}
\usepackage{braket}
\usepackage{outlines}
\usepackage{enumitem}
\usepackage{soul}
\usepackage{color}

\usepackage{bm} 

\usepackage[utf8]{inputenc}
\usepackage{siunitx,booktabs}
\usepackage{multirow}
\usepackage{amssymb}

\usepackage{relsize}

\usepackage{soul}
\usepackage{tabularx}

\usepackage{amsmath}
\usepackage{amssymb}
\usepackage{bbm}
\usepackage{braket}
\usepackage{xcolor}

\usepackage{comment}
\usepackage{pifont}

\begin{document}

\title{Theory of excitonic order in kagome metals ScV$_6$Sn$_6$ and LuNb$_6$Sn$_6$}
 
\author{Julian Ingham}
\email[]{ji2322@columbia.edu}
\affiliation{Department of Physics, Columbia University, New York, NY, 10027, USA}

\author{Armando Consiglio}
\affiliation{Istituto Officina dei Materiali, Consiglio Nazionale delle Ricerche, Trieste I-34149, Italy}

\author{Domenico di Sante}
\affiliation{Department of Physics and Astronomy, University of Bologna, 40127 Bologna, Italy}

\author{Ronny Thomale}
\email[]{ronny.thomale@uni-wuerzburg.de}
\affiliation{Institute for Theoretical Physics and Astrophysics, University of Würzburg, D-97074 Würzburg, Germany}
\affiliation{Department of Physics and Quantum Centers in Diamond and Emerging Materials (QuCenDiEM) Group, Indian Institute of Technology Madras, Chennai, India}

\author{Harley D.~Scammell}
\affiliation{School of Mathematical and Physical Sciences, University of Technology Sydney, Ultimo, NSW 2007, Australia}

\date{\today}

\begin{abstract}

We argue that kagome metals can feature an excitonic condensate of unconventional nature. Studying the recently discovered variants ScV$_6$Sn$_6$ and LuNb$_6$Sn$_6$ we identify electron and hole pockets due to a pair of van Hove singularities (vHS) close to the Fermi level, with an approximate spectral particle-hole symmetry. A significant fraction of the Fermi level density of states away from the vHS is removed by the onset of high temperature charge density wave order, and makes the bands more two-dimensional, setting the stage for the formation of excitons. We develop a two-orbital minimal tight-binding model of these materials which captures these features along with the sublattice support of the wavefunctions, and find $s$- or $d$-wave excitons depending on interaction parameters -- the latter of which exhibits either charge nematicity or time-reversal symmetry breaking (TRSB) depending on strain, offering an explanation of recent STM and transport experiments. The presence of particle- and hole-type vHS, and the associated excitonic resonance, may be a common thread to understanding nematicity and TRSB in kagome metals.
\end{abstract}

\maketitle

{\it Introduction.} The kagome pattern -- consisting of a network of corner sharing triangles -- has long been seen as a promising motif in the search for materials hosting novel physics \cite{Johnston1990,Yu2012,Wen2010,Kiesel2013,Kiesel2012,Li2022}. In the last decade, increasing attention has been directed towards kagome metals, offering a rich kinematic phenomenology involving Dirac points, flat bands and van Hove singularities (vHS) as a function of Fermi level \cite{wang2023quantum,yin2022topological}. As compared to previous studies of correlated electrons in hexagonal systems such as the triangular lattice or the honeycomb lattice, the kagome lattice stands out in that vHS are naturally prone to appear in the vicinity of the Fermi level. 

As a common thread to electronic orders discovered thus far in kagome metals, such as the V-based $A$V$_3$Sb$_5$ ($A$=K,Rb,Cs) \cite{Ortiz2020, wilson2024a, jiang2023kagome}, or Ti-based $A$Ti$_3$Bi$_5$ \cite{yang2022superconductivity, liu2023tunable}, a high-temperature charge order appears, followed by a superconducting transition at lower temperature. Compound-dependant differentiations proliferate with regard to the precise nature of charge order and superconductivity -- ranging from rotational and time-reversal symmetry breaking, as well as the possibility of pair density wave or a unit-cell modulated superconducting pairing \cite{Li2021b, nie2022charge, li2023electronic, jiang2023flat, Xu2022, Jiang2021, Chen2021}.

\begin{figure}[bt]
\centering
\includegraphics[width = 0.95\columnwidth]{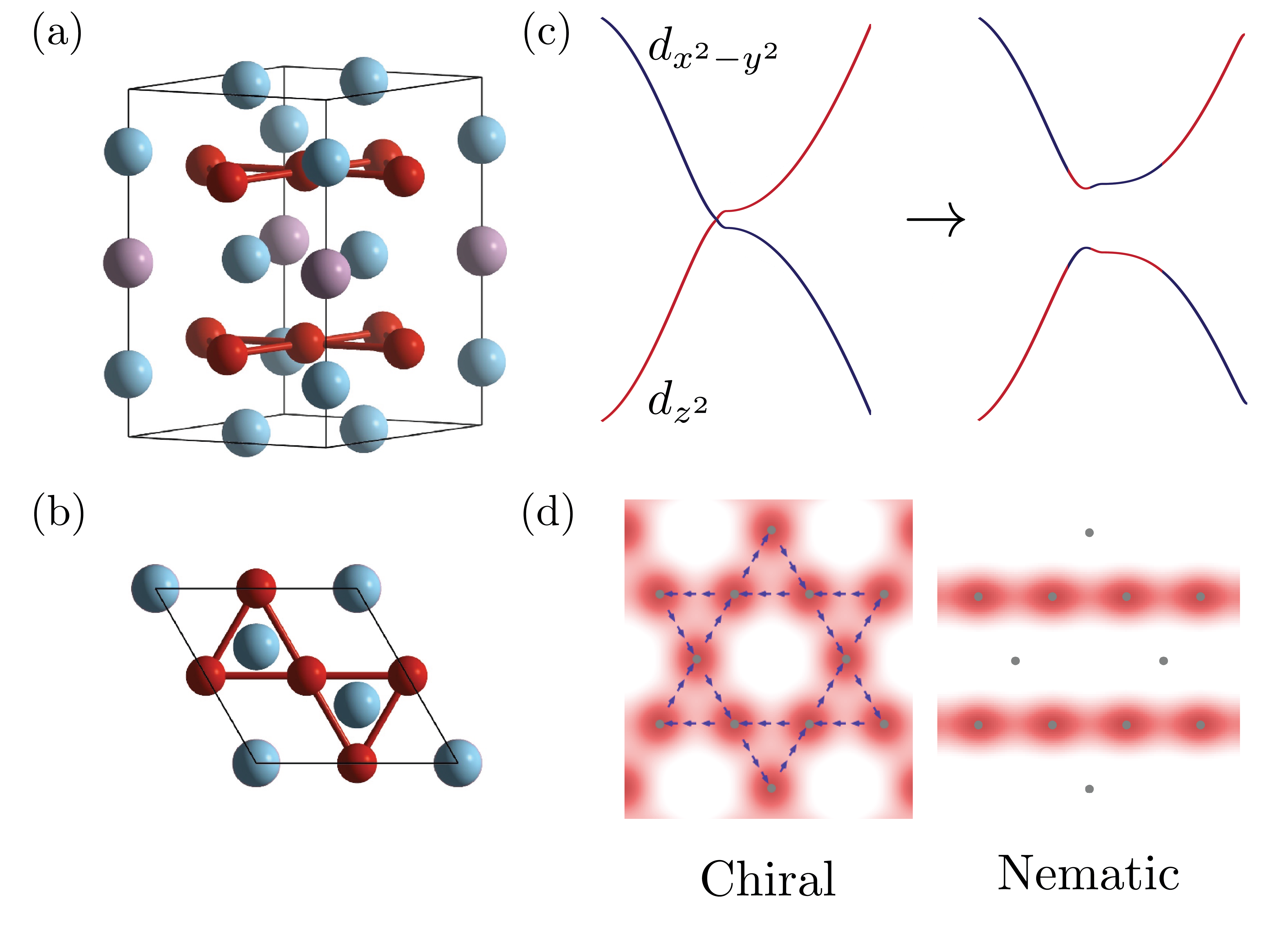}	
\vspace{-0.2cm}
\caption{\textbf{Excitonic orbital order in 166 kagome metals.} The crystal structure of ScV$_6$Sn$_6$: (a) standard 
and (b) top-view of the pristine unit cell (V = red, Sc = violet, Sn = blue). (c) The presence of oppositely dispersing vHS results in a weak coupling instability towards an excitonic state which hybridizes the two bands, producing either nematic order (d, right) in which the density of states (red) breaks rotational symmetry, or flux order (d, left) with rotationally symmetric density of states but with spontaneous loop currents (arrows).}
\label{f:structure}
\vspace{-0.7cm}
\end{figure}

Excitons are emergent bound states formed in insulators and semiconductors between a electron-like and hole-like carrier, where the binding energy derives from Coulomb interactions. The constraints on the band structure of a crystal in which such excitons may form, and be experimentally resolvable, are manifold. An idealized toy model comprises a valence and conduction band of opposite curvature, and neglects exciton decay via additional bands or other midgap states. As with any other two-particle state, the excitonic particle-hole pair relates to an irreducible point group representation (irrep) and inherits its specific form from the spin and orbital nature of the underlying electronic degrees of freedom. If the gap is small, or even zero in the case of a semimetal, the exciton energy can become negative, triggering Bose condensation and a symmetry-breaking ground state \cite{keldysh1965possible, jerome1967excitonic, halperin1968excitonic}.

In this Letter, we develop a theory of excitonic resonances in kagome metals. Exploiting the abundance of vHS around pristine filling, we consider the case where opposite concavity vHS occur at a similar distance from the Fermi level. A properly chosen charge order can effectively convert a kagome metal to a `kagome semimetal' at low energies, by gapping out a large fraction of complementary Fermi level density of states while leaving the vHS largely unspoiled. We show that this holds for recently discovered variants ScV$_6$Sn$_6$ and LuNb$_6$Sn$_6$ \cite{Arachchige2022, ortiz2025stability}, in contrast to $A$V$_3$Sb$_5$ in which the CDW develops within the same orbitals as a possible excitonic state \cite{Scammell2023,ingham2025vestigial}. Developing a minimal model, we find $s$-wave and $d$-wave excitonic states as a function of interaction parameters, the latter of which offers an explanation of the nematic state recently seen in STM \cite{Jiang2023hove}. The excitonic scenario we develop for ScV$_6$Sn$_6$ provides a theoretical reconciliation of several experimental facts, and suggests novel signatures in optical spectroscopy.

{\it Ab initio theory and minimal model.} ScV$_6$Sn$_6$ is a paramagnetic metal belonging to a large family of related 166 compounds \cite{Jiang2023,Hu2023b,DiSante2023} of the form $AM_6 X_6$ (e.g. $A$=Tb,Ho,Sc; $M$=V,Cr,Fe; $X$=Si,Ge,Sn) and consists of a kagome bilayer structure comprising a triangular lattice of scandium interposed with a honeycomb lattice of tin, two triangular lattices and a honeycomb lattice of tin, and two kagome lattices of vanadium atoms (Fig. \ref{f:structure}). The high temperature phase belongs to the space group $P6/mmm$ (SG 191) \cite{Arachchige2022, Gu2023,Hu2023c}, and is reduced to R$\bar{3}m$ (SG 166) \footnote{The CDW structure originally reported in Ref. \cite{Arachchige2022} is weakly inversion symmetry breaking, corresponding to R$32$ (SG 155, c.f. \cite{HINUMA2017140, togo2024spglib}).} by a first-order transition to a $(\tfrac{1}{3},\tfrac{1}{3},\tfrac{1}{3})$ CDW at 92 K. Above $T_{CDW}$, fluctuations at $(\tfrac{1}{3},\tfrac{1}{3},\tfrac{1}{2})$ are observed but do not order \cite{Pokharel2023,Cao2023,Korshunov2023}, suggesting a competing CDW order also seen in first principles calculations \cite{Tan2023,Liu2023,Subedi2023}.

The CDW is primarily concentrated on the Sn and Sc layers, and is largely absent on the vanadium kagome layers, as is known from STM \cite{Hu2023c} and the lack of a gap observed on the vanadium bands \cite{Hu2023,Kim2023,Kang2023,Cheng2024,arpes,kundu2024low}. This can be understood physically -- the CDW  originates from the softening of an out-of-plane phonon mode corresponding to oscillations of the Sn and Sc atoms \cite{Tuniz2023}, which behave roughly as a set of weakly coupled one-dimensional chains. Substitution of Sc with larger rare earth ions inhibit the Sn-Sn oscillations, explaining the doping suppression of the CDW \cite{Meier2023}. In contrast to the Sb $p$-orbitals in $A$V$_3$Sb$_5$, the Sn $p_z$ orbitals contribute much more weakly to the Fermi surface -- via trigonal rather than hexagonal Sn -- hybridizing with the V atoms to produce $k_z$ dispersion \cite{Korshunov2023} rather than contributing significantly to the in-plane dispersion, which is primarily derived from the V $d$-orbitals.



The CDW is more rapidly suppressed with applied pressure or doping compared to $A$V$_3$Sb$_5$, though superconductivity has not been seen in this material, at any pressure or doping explored to date \cite{Meier2023,transport2}. This suggests a further comparison with $A$V$_3$Sb$_5$: one viewpoint holds that superconductivity is driven by the Sb $\Gamma$-pocket; an analogous Sn band is absent in the 166 family.


\begin{figure}[bt]
\includegraphics[width = 0.95\columnwidth, trim={4.6cm 14cm 7.5cm 4.2cm},clip]{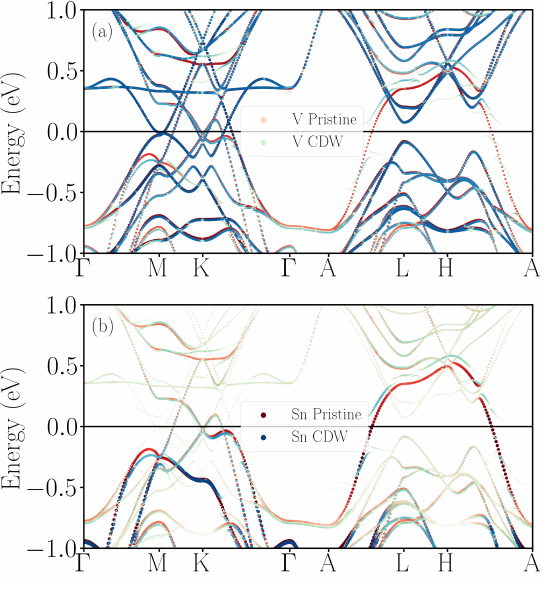}		
\vspace{-0.4 cm}
\caption{\textbf{Ab initio bandstructure in the pristine and CDW phase of ScV$_6$Sn$_6$.} DFT bands above (red) and below (green) the CDW transition, with V and Sn orbital weights. For better visualization, the Sn weights are scaled by a factor four times greater than the V weights. The band crossing the Fermi level in the $k_z = 1/2$ plane derives from the hybridization between $p_z$ orbitals of Sn atoms and V atoms; the associated spectral weight is gapped out by the CDW, producing a more two-dimensional electronic structure.}
\label{f:dft_cdw}
\vspace{-0.55 cm}
\end{figure}

\begin{figure*}[t!]
\centering
\includegraphics[width = 1\textwidth, trim={1.4cm 6.5cm 0.5cm 4.55cm},clip]{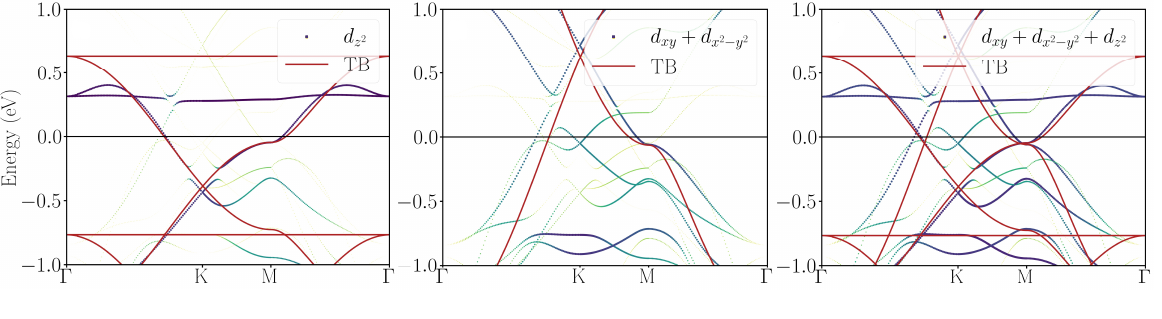}
\vspace{-1.0 cm}
\caption{\textbf{Minimal model compared with DFT.} Comparison between orbitally-projected DFT bands, in the pristine phase case, and the two-orbital tight-binding model, Eq. \eqref{Htb}. Orbital weight varies from dark blue (large) to yellow (small).}
\label{f:ab_init}
    \vspace{-0.35cm}
\end{figure*}

Recently, a sister compound LuNb$_6$Sn$_6$ has been discovered \cite{ortiz2025stability} which exhibits very similar phenomenology, including the competing CDW orders \cite{wang2025formation}, orbital content and bandstructure. Many of our predictions apply similarly to this compound, and so we shall focus our attention on ScV$_6$Sn$_6$, presenting relevant details on LuNb$_6$Sn$_6$ in the Supplementary Material (SM).

These materials are significantly more three dimensional than the $A$V$_3$Sb$_5$ series, due to the hybridization with the out-of-plane Sn orbitals -- as reflected by the three-dimensional character of the dispersion shown in Fig. \ref{f:dft_cdw}. This is also borne out in the lower transport anisotropy $\rho_c/\rho_{ab} \approx 5$ \cite{Mozaffari2023,DeStefano2023} and relative difficulty of cleaving samples \cite{Kang2023}. Yet, the spectral weight associated to the $p$-orbitals of the Sn atoms is gapped out upon the onset of the CDW, resulting in a significantly more two-dimensional dispersion, as well as pushing the two vHS closer to the Fermi level, Fig. \ref{f:dft_cdw}, also reflected in the Fermi surface plots shown in the SM. As the CDW primarily affects the Sn and Sc/Lu atoms, the V-based kagome bandstructure remains intact.

\begin{figure}[b!]
\includegraphics[width = 1\columnwidth, trim={0.1cm 9.5cm 0.40cm 5.8cm},clip]{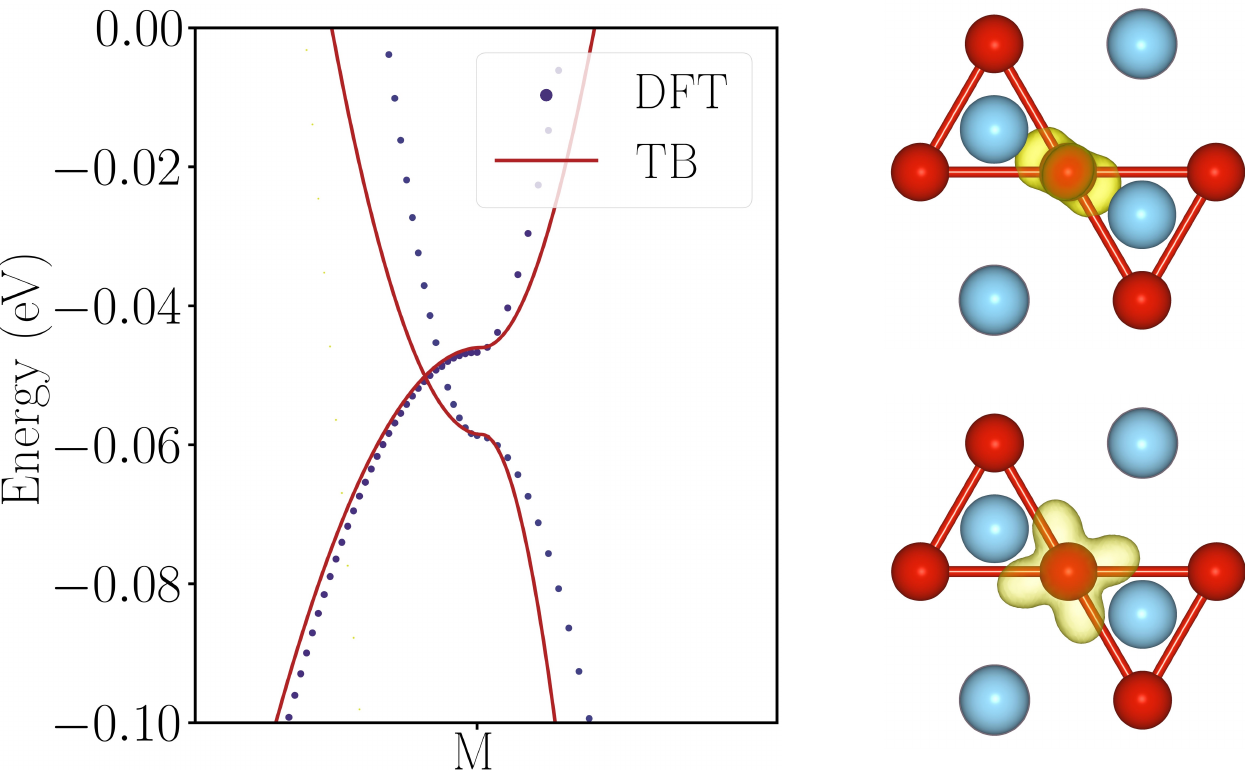}	
\vspace{-0.5cm}
\caption{\textbf{Pure type sublattice composition of the two vHS near the Fermi level.} (Left) Close up of the vHS arising from the two orbital model, compared with DFT. The upper vHS mainly has V $d_{z^2}$ character, while the lower vHS mainly has V $d_{xy} + d_{x^2-y^2}$ character. Right: Top-view of the partial charge density near $M$ for the upper vHS with $d_{z^2}$ character (top) and lower vHS with $d_{x^2-y^2}/d_{xy}$ character (bottom). In agreement with the $p$-type nature of the vHS, only one of three sites contributes.}
\label{f:ptype}
\vspace{-0.3 cm}
\end{figure}

Hence, we can construct a minimal model of ScV$_6$Sn$_6$ and LuNb$_6$Sn$_6$ inside the CDW phase consisting of two effective $d$ orbitals per site on a kagome bilayer, representing the twelve bands predominantly composed of the vanadium $d_{x^2-y^2}$/$d_{xy}$ and $d_{z^2}$ orbitals respectively, corresponding to the $A_{2g}$ and $A_{1g}$ representations of the site symmetry group respectively. Since the dispersion arising from vanadium orbitals varies minimally along the $k_z$ direction inside the CDW phase, we work with a model that neglects the hoppings along the $z$ direction. Moving from layer basis to the basis of mirror eigenstates $(|\text{top}\rangle\pm |\text{bottom}\rangle)/\sqrt{2}$, the bands decouple into four sets of kagome bands with distinct mirror eigenvalues; projecting onto the two which cross the Fermi level, we obtain
    \begin{align}
        \mathcal{H} = -\sum_{i,\sigma\neq\sigma'} t_{\alpha} \psi^\dag_{i\alpha\sigma} \psi_{j\alpha\sigma'} + \sum_{i} \varepsilon_\alpha \psi^\dag_{i\alpha\sigma} \psi_{i\alpha\sigma} 
        \label{Htb}
    \end{align}
where $\alpha=z,x$ indexes the two orbitals, $\sigma$ sublattice, $t_x=0.69$ eV is the in-plane hopping between the $d_{xy}$-like orbitals, $t_z=-0.34$ eV is the in-plane hopping between the $d_{z^2}$ orbitals, and additionally, the two sets of orbitals possess different onsite energies, $\varepsilon_z = -0.77$ eV and $\varepsilon_x = 1.35$ eV. Comparison between DFT and tight-binding in Fig. \ref{f:ab_init} indicates good agreement \footnote{Ref. \cite{Jiang2023} has also discussed tight-binding minimal models for 166 and the related 1:1 compounds. The models considered there apply to the pristine phase including the Sn $p$ orbitals, whereas our present model neglects the role of the $p$ orbitals, which we justify with resort to the effect of the CDW on the Sn spectral weight (Fig. \ref{f:dft_cdw}).}. Our model neglects the bands which cross the Fermi level near $K$ (Fig. \ref{f:dft_cdw}), but these bands are poorly nested with the bands at $M$, and so do not contribute to the Fermi surface instabilities we describe. DFT and tight-binding fits for the two-orbital model of (Lu,Ho,Tb)Nb$_6$Sn$_6$ are given in the SM.

A key purpose of our minimal model is to accurately characterize the two vHS located close to the Fermi level, producing electron-like and hole-like Fermi surfaces with an approximate emergent particle-hole symmetry, consistent with prior ab initio studies \cite{Tuniz2023} and ARPES \cite{arpes,kundu2024low} which observe two oppositely-dispersing vHS. The kagome tight-binding model exhibits a novel wavefunction structure near the vHS -- referred to as the sublattice interference effect. Namely, the wavefunctions possess support either on one sublattice (pure, or $p$-type, sublattice support) or two (mixed, or $m$-type, sublattice support). We note importantly that the two vHS near the Fermi level are both $p$-type -- as illustrated in Fig. \ref{f:ptype}. The plot of the charge density obtained from DFT shows that near a given $M$-point, the wavefunction of both bands is concentrated entirely on one sublattice.  An accurate reflection of the sublattice support plays an important role in our analysis of interaction effects.

\begin{figure*}[t!]
    \centering
    \includegraphics[width=0.99\textwidth]{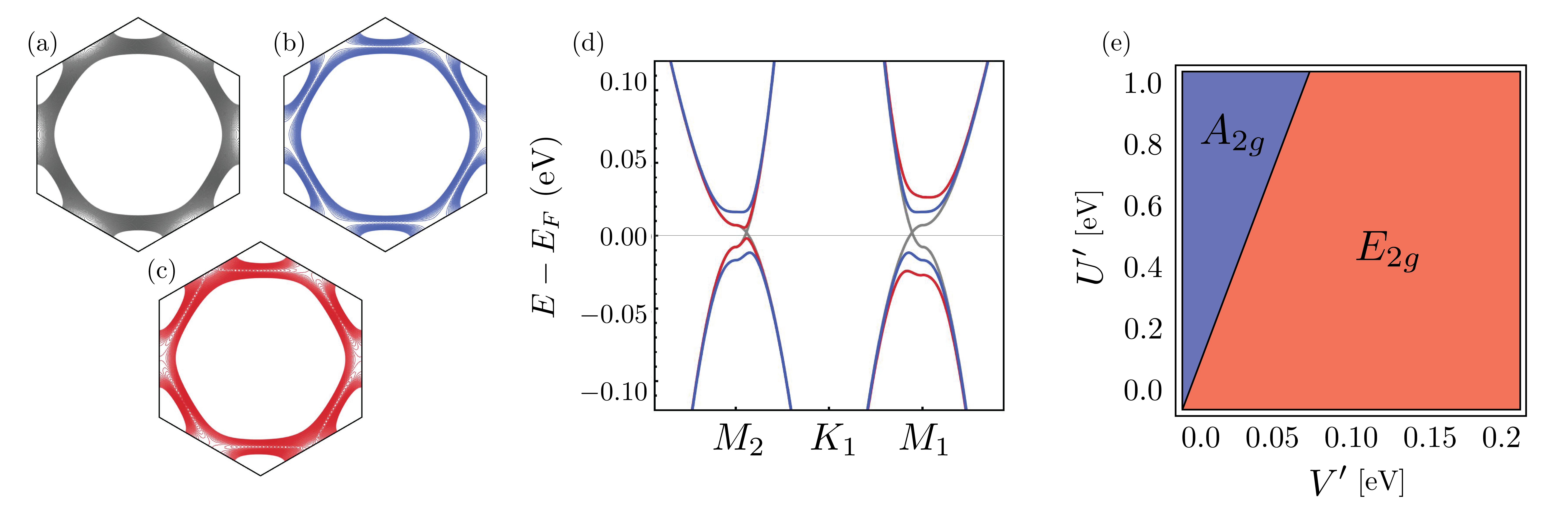}
    \vspace{-0.5cm}
    \caption{\textbf{Mean-field excitonic states.} Contour plots of the minimal model (a) without excitonic order, and in the presence of (b) $A_{1g}$ excitonic order, and (c) one component of $E_{2g}$ excitonic order. Also shown (d) band structure cut with same colour coding. (e) Phase diagram as a function of onsite and nearest-neighbour inter-orbital Hubbard parameters.}
    \label{solutions}
    \vspace{-0.4cm}
\end{figure*}

{\it Excitonic order.} The emergence of excitons can be understood as a two-particle pairing problem: we first perform a particle-hole transformation, taking the valence electron with negative vHS concavity to a hole with positive concavity. The conduction electron and hole then have same dispersion but opposite charge. The bare Coulomb repulsion then becomes an effective attraction, driving the formation of bound states. The questions we address here are: ($i$) can these bound states condense, and ($ii$) in which spatial irrep do they condense. 

On point ($i$): the two vHS cross each other, forming a `kagome semimetal'; if this band-crossing lies directly at the Fermi level, then for arbitrarily small dimensionless Coulomb interaction $g$, there exists a non-zero critical temperature at which excitons condense \footnote{The origin of weak-coupling instability towards exciton condensation can be understood as `zero wavevector nesting': as is well known, when the Fermi surface possesses the nesting property $\varepsilon(\bm{k}) = -\varepsilon(\bm{k} + \bm{Q})$ for some wavevector $\bm{Q}$, the Fermi surface becomes unstable towards the formation of a spin or charge density wave $ \langle c^\dag_{\bm k+\bm Q} c_{\bm k} \rangle \neq 0 $. In the presence of two Fermi surfaces $c$ and $d$ with $\varepsilon_c(\bm{k}) = -\varepsilon_d(\bm{k})$, one finds the nesting condition satisfied with $\bm{Q}=0$, indicating an instability towards a state which hybridizes the two Fermi surfaces $ \langle c^\dag_{\bm k} d_{\bm k} \rangle \neq 0 $.}. If the vHS cross away from Fermi level, as in ScV$_6$Sn$_6$ and LuNb$_6$Sn$_6$, then there is an energy cost $\delta \varepsilon$ to create particle-hole pairs, which then appear as gapped states with some lifetime. Upon including interactions, the exciton has a total energy $\delta\varepsilon-\varepsilon_b$, where $\varepsilon_b$ is the binding energy. For $\varepsilon_b>\delta\varepsilon$, condensation occurs; this requires a minimum coupling strength $g^*$ such that $\delta\varepsilon = \Lambda e^{-1/\sqrt{g^*}}$, where $\Lambda$ is an energy cutoff \footnote{The $e^{-1/\sqrt{g}}$ dependence, rather than the $e^{-1/g}$ typical of BCS-like instabilities, appears due to the enhanced density of states near a vHS.}. For ScV$_6$Sn$_6$, the relatively small $\delta\varepsilon \approx 20$ meV \cite{arpes} suggests a tendency towards condensation \footnote{Note that DFT predicts $\delta \varepsilon \sim$ 50 meV, c.f. Fig. \ref{f:ptype}, overestimating the value of $\delta\varepsilon$ seen in experiment.}.


To address point ($ii$), we present an analysis of excitonic order at mean field level in a minimal interacting model. The interactions which contribute to the mean-field equations are those which couple electrons and holes, which in our model correspond to the $d_{x^2-y^2}/d_{xy}$ and $d_{z^2}$ orbitals. In the SM, we present a larger set of onsite and nearest-neighbour inter-orbital Hubbard interactions, but here we present a minimal case invoking an onsite inter-orbital $U'$, and nearest neighbour orbital exchange interaction $V'$, \begin{gather}
\notag H_\text{int} = \sum_{i,\sigma \sigma',\alpha\neq\alpha'}  U'\delta_{\sigma\sigma'}(\psi^\dag_{i\alpha\sigma}\psi_{i\alpha\sigma})(\psi^\dag_{i\alpha'\sigma}\psi_{i\alpha'\sigma})\\
\label{Hint}
    +  V'(1-\delta_{\sigma\sigma'})(\psi^\dag_{i\alpha\sigma}\psi_{i\alpha'\sigma})(\psi^\dag_{i\alpha'\sigma'}\psi_{i\alpha\sigma'})
\end{gather}
with $\alpha$ and $\sigma$ the orbital and sublattice indices, and $U',V'>0$.
We denote the creation operators for electron and hole-like fermions in band basis as $c^\dag_{\bm k}$ and $d^\dag_{\bm k}$, related to the sublattice $\sigma$ basis via the unitary transformation $u_{\bm k, +, \sigma}c^\dag_{\bm k} = \psi^\dag_{{\bm k},+,\sigma}$ and $u_{\bm k, -, \sigma}d^\dag_{\bm k} = \psi^\dag_{{\bm k},-,\sigma}$. Defining the vertex form factors $F^{\mu\nu}_{\bm k_1,\bm k_2; \sigma_1,\sigma_2} = u^*_{\bm k_1, \mu, \sigma_1}u_{\bm k_2, \nu, \sigma_2}$, and performing the mean-field decomposition for $\Phi_{\bm k}=\langle c^\dag_{\bm k}d_{\bm k} \rangle$ we arrive at the excitonic gap equation,
 \begin{gather}
    \notag \Phi_{\bm k_1}
    =\!\sum_{\bm k_2, \sigma, \sigma'}  \left[U'\delta_{\sigma,\sigma'}-2V'(1-\delta_{\sigma,\sigma'})\right]\\
    \times|F_{\bm k_1, \bm k_2; \sigma\sigma'}|^2 \, \Pi_{\bm k_2} \Phi_{\bm k_2}
    \label{verteq_main} 
\end{gather}
with $\Pi_{\bm k}$ the particle-hole susceptibility (see the SM).
Reaslistic values of $\delta \varepsilon$ have negligible effect on the phase diagram, since $\delta \varepsilon \ll U', V'$ \footnote{We leave first-principles estimates of $U'$ and $V'$ to future work, but note that cRPA studies of 135 kagome metals with similar orbital content find $U',V'\sim$1-2 eV \cite{Wu2021}, and so we produce a phase diagram for a broad window of coupling constants with this order of magnitude.}, and so we shift the vHS to lie exactly at the Fermi level and solve numerically for $\Phi_{\bm k}$ in the $s$- and $d$-wave like irreps, $\Gamma_\Phi= \{A_{1g}, E_{2g}\}$. Note that due to the Perron-Frobenius theorem, the lowest energy bound state of a purely attractive interaction potential will always be the $A_{1g}$ state -- a theorem well-noted in the context of superconducting instabilities \cite{scheurer2024mechanism}. Similarly, a purely attractive potential, such as that arising from $U'$, will always favor $A_{1g}$ excitons. The orbital exchange term $-2V'<0$  generates repulsion for excitons on different sublattices -- and hence at different vHS, due to the $p$-type sublattice support -- facilitating the appearance of $E_{2g}$ order. Undoing the particle-hole transformation, one finds that $V'$ acts as an effective attractive interaction due to the unusual sublattice structure of the kagome lattice. Our results therefore reflect a subtle and new mechanism for nematicity in kagome systems, derived from a competition between onsite and neighbouring Coulomb interactions.

Note that our use of the Perron-Frobenius theorem applies to other particle-hole instabilities, including ferromagnetism; for this reason, most ordinary ferromagnets are $s$-wave, possessing trivial spatial structure. In recent years, great attention has been devoted to altermagnets, a kind of non $s$-wave ferromagnet \cite{smejkal2022beyond, smejkal2022emerging}. Our arguments imply that a purely repulsive Coulomb interaction cannot induce an altermagnetic phase via Fermi surface instability -- some effective attraction is necessary, either from phonons or a non trivial sublattice structure, placing useful constraints on models of altermagnetism.

The symmetry representation $\Gamma$ of the excitonic state is a product of the irreps of the $c$ and $d$ bands with the transformation of $\Phi_{\bm k}$ in momentum space, $\Gamma = \Gamma_{\Phi}\otimes\Gamma_{c}\otimes\Gamma_{d}$. Interestingly, since the $c$ and $d$ bands possess different symmetries, this results in the $A_{1g}$ solution for $\Phi$ corresponding to an $A_{2g}$ order parameter, transforming non-trivially under mirror when taking into account the symmetry properties of the orbitals \footnote{The symmetry behaviour of the $E_{2g}$ state is unchanged upon taking into account the band representations.}.

We plot the resulting Fermi surface for both order parameters in Fig. \ref{solutions}a-c, alongside the associated bandstructure along a high-symmetry cut. For the $A_{2g}$ solution, the two vHS are gapped out completely, while for a single $E_{2g}$ solution the vHS remain gapless at one of the $M$ points.  We present the full momentum dependence of the eigenvectors in the SM. We find the excitonic state is very sensitive to the orbital exchange interaction $V'$ in our simplified model, developing in the $E_{2g}$ channel for small $V'/U'\gtrsim 0.1$, as shown in Fig. \ref{solutions}e. In the SM, we present additional parameter space which illustrate that other inter-orbital Hubbard parameters can make the $A_{2g}$ state more competitive.

The excitonic state breaks an approximate U(1) symmetry associated to the difference in particle number of electrons and holes. Since this is not a microscopically conserved quantity, we expect small explicit symmetry breaking terms to pin the complex phase of this order parameter at low temperatures -- either due to interorbital hopping terms, which are known to be parametrically weak \cite{Wu2021}, or `orbital umklapp' interactions. Such effects break the emergent U(1) to a residual $\mathbb{Z}_2$, producing excitonic states that are either purely real or imaginary \footnote{This is akin to the appearance of purely real and imaginary CDW states -- rCDW and iCDW -- in patch models, which arises due to conventional umklapp interactions between the different vHS breaking an emergent ``patch conservation'' symmetry, which would otherwise produce complex CDW orders with an arbitrary U(1) phase.}. A purely imaginary $A_{2g}$ state couples directly to magnetic field -- this state has the symmetries of a `flux phase', and unusually, breaks time-reversal symmetry despite being a single component order parameter.

The $E_{2g}$ solutions to the gap equation are twofold degenerate, $\varphi_{1,2}$. The ground state may consist of a chiral combination of these two states $\varphi_1 + i\varphi_2$, or else a nematic real combination e.g. $\varphi_1+\varphi_2$. Magnetotransport and $\mu$SR see evidence of ``soft orbital ferromagnetism'' -- weak signatures of time-reversal symmetry breaking which are enhanced in an applied field \cite{musr,transport}; the $A_{2g}$ flux and chiral $d$-wave orders are both possible candidates for time-reversal symmetry breaking, which may be enhanced over the nematic combination by an applied field. While generically a mean-field computation of the Landau-Ginzburg free energy favours the chiral combination $\varphi_1+i\varphi_2$, strained samples can produce nematic order \cite{sigrist1991}. We also note that there are cases in which variational methods find strong Coulomb repulsion results in a nodal rather than fully-gapped chiral state, despite the expectations of mean-field theory \cite{Grover2010, Guerci2024}.

{\it Discussion.} In this work we have developed a minimal model for 166 kagome metals in the $\sqrt{3}\times \sqrt{3}$ CDW phase, and proposed that this system hosts excitonic order, discussing a new mechanism for non $s$-wave orders and offering an explanation of recent experiments. We find both a fully gapped phase which preserves  time-reversal symmetry, and a nematic state which breaks rotational symmetry and selectively gaps the vHS -- both states being possibly relevant to experiment. Interestingly, our prediction of nematicity and possible flux order relied solely on the nature of the bandstructure and orbital content, which we have shown to apply also to LuNb$_6$Sn$_6$; we therefore predict the observation of nematicity, and unusual transport signatures in this sister compound.

STM studies on kagome surface terminations of ScV$_6$Sn$_6$ observe nematic order around $70$ K -- clearly distinguished from the CDW, which is absent on the kagome surface \cite{Jiang2023hove}. Quasiparticle interference imaging observes a gap at the vHS at two of the three sets of $M$-points at this temperature; the appearance of a gap at the crossing between two oppositely-dispersing bands is a direct signature of an excitonic state, distinct from signatures produced by other nematic states \cite{nab2024pomeranchuk}. The fact that the nematic state affects the bandstructure at $M$ most strongly indicates that the crucial bands are those in our two orbital model, which neglects the bands near $K$. The excitonic state may also be related to the transport signatures of a second transition within the CDW phase \cite{Mozaffari2023} -- near which nematic features have recently been seen \cite{farhang2025discovery} -- and weak spectral features seen near the vanadium vHS \cite{Chen2023,Yang2024unveiling}. 

We lastly discuss more direct experimental probes of excitonic physics. Above the temperature at which excitons condense, they appear as excitations of the system with an energy that gradually reduces upon approaching the critical temperature. Excitons with zero center of mass momentum are referred to as `bright', and `dark' otherwise; the translationally-invariant nature of our excitonic order parameter means the associated excitons are bright. Optical spectroscopy is a powerful tool to probe for such resonances; observing these excitations requires the exciton energy to be sufficiently large and to exceed the linewidth induced by various scattering processes.

Our related theoretical work in Ref. \cite{Scammell2023} predicted the appearance of such states in kagome metal CsV$_3$Sb$_5$, where they are expected to appear as bright resonances inside the CDW gap, due to coexistence of excitons with CDW of the same orbital content. In the case of ScV$_6$Sn$_6$ and LuNb$_6$Sn$_6$, excitons form in a parent bandstructure which is semimetallic, which makes observation of these resonances more challenging as they are expected to be broad due to coupling with the electronic continuum. Excitons in the former situation are similar to those in transition metal dichalcogenides \cite{wang2018excitons}, whereas those in the latter are analogous to those in candidate excitonic insulator Ta$_2$NiSe$_5$, a zero-gap semiconductor \cite{lu2017zero}.

In both the cases of $A$V$_3$Sb$_5$ and ScV$_6$Sn$_6$/LuNb$_6$Sn$_6$, the coexisting metallic bands which do not form excitons comprise different orbital content, suggesting that matrix element effects may suppress the exciton scattering rate and permit a feasible resolution; we leave a theoretical analysis of this possibility to future work, but suggest experimental investigation is the best approach to answer these questions.

\section*{Acknowledgments}
J.I. thanks Riccardo Comin, Zurab Guguchia, Xiong Huang, Yu-Xiao Jiang, Chenhao Jin, Asish Kundu, William Meier, and Abhay Pasupathy for helpful discussions and comments. J.I. is supported by NSF Career Award No. DMR-2340394. A.C. and R.T. acknowledge useful conversations with Matteo Dürrnagel, Werner Hanke, Steve Kivelson, and Riccardo Sorbello.  A.C. acknowledges support from PNRR MUR project PE0000023-NQSTI. A.C., D.D.S. and R.T. acknowledge the Gauss Centre for Supercomputing e.V. (https://www.gauss-centre.eu) for funding this project by providing computing time on the GCS Supercomputer SuperMUC-NG at Leibniz Supercomputing Centre (https://www.lrz.de).

\let\oldaddcontentsline\addcontentsline
\renewcommand{\addcontentsline}[3]{}

\newpage \widetext \newpage

\begin{center}
\textbf{\large Supplementary Material}
\end{center}

\setcounter{equation}{0}
\setcounter{table}{0}
\setcounter{section}{0}
\setcounter{figure}{0}
\renewcommand{\theequation}{S\arabic{equation}}
\renewcommand{\thefigure}{S\arabic{figure}}
\renewcommand{\thesection}{S\arabic{section}}

\renewcommand{\bibnumfmt}[1]{[S#1]}
\renewcommand{\citenumfont}[1]{S#1}
\section{Bandstructure details}

Density functional theory calculations were performed with the Vienna ab initio Simulation Package (VASP) \cite{PhysRevB.59.1758, PhysRevB.54.11169}, using the projector augmented wave (PAW) method \cite{PhysRevB.50.17953}. Exchange and correlation effects have been included at the generalized gradient approximation (GGA) level \cite{PhysRevB.46.6671, PhysRevA.38.3098, PhysRevB.28.1809} within the Perdew-Burke-Ernzerhof (PBE) approach \cite{PhysRevLett.77.3865}. 
The plane-wave cutoff used for the truncation of the basis set is 500 eV, both for the pristine and the CDW unit cells. In both cases the relaxation of the electronic and ionic degrees of freedom was considered converged when the output difference between two steps was equal or smaller than $1\times10^{-6}$ eV and $1\times10^{-6}$ eV/\AA, respectively. The number of $\mathbf{k}$-points used for the pristine unit cell is 15$\times$15$\times$9, while the CDW supercell it is 9$\times$9$\times$3. Partial occupancies have been determined via Gaussian smearing with a width of 0.01 eV.
Spin-orbit coupling has not been considered in this work. DFT band structures have been visualized using the VASPKIT postprocessing tool \cite{WANG2021108033}, while VESTA \cite{Mommako5060} has been used to visualize the crystal structures and the charge densities.

\subsection{Fermi surface for ScV$_6$Sn$_6$}
To complement our plots of the bandstructure in the main text, we present plots of the Fermi surface in Fig. \ref{f:ab_init}, which make more explicit the increased two-dimensionality inside the CDW phase.

\begin{figure*}[h!]
\centering
\includegraphics[width = 0.85\textwidth, trim={0.5cm 19cm 0.8cm 2.2cm},clip]{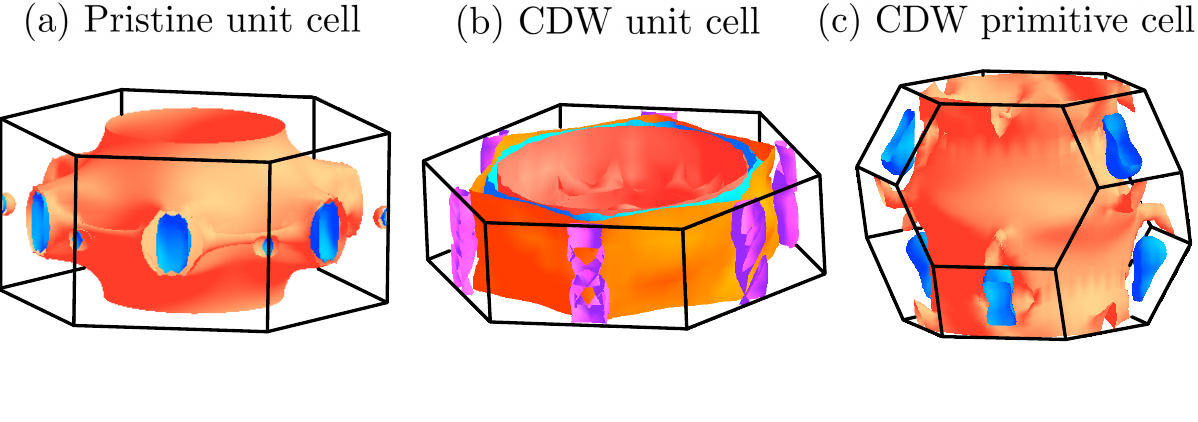}
\vspace{-1.0 cm}
\caption{\textbf{Fermi surfaces for ScV$_6$Sn$_6$.} Fermi surface in the (a) pristine phase above $T_{CDW}$, and beneath $T_{CDW}$ in the (b) unit cell, and (c) primative cell.}
\label{f:fermi_surface}
\end{figure*}

\subsection{Electronic structure of LuNb$_6$Sn$_6$}

\begin{figure*}[h!]
\centering
\includegraphics[width = 0.95\textwidth, trim={0cm 19.5cm 0cm 2.5cm},clip]{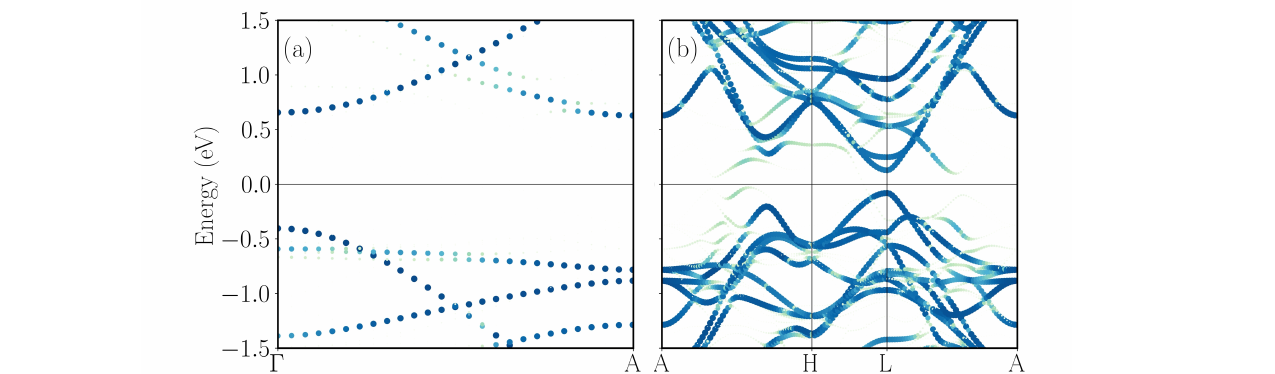}	
\caption{{\textbf{Influence of CDW order on the out-of-plane dispersion in LuNb$_6$Sn$_6$.} (a) Electronic structure of LuNb$_6$Sn$_6$ in the CDW phase, along the $\Gamma$-A path. (b) Electronic structure of LuNb$_6$Sn$_6$ in the CDW phase, along the A-H-L-A path. Note the opening of the Nb band around the Fermi level, similarly to the V band of ScV$_6$Sn$_6$ in the CDW configuration.}}
\label{f:kzonehalf_CDW}
\end{figure*}

\begin{figure*}[h!]
\centering
\includegraphics[width = 0.945\textwidth, trim={1.1cm 58cm 1.2cm 3.5cm},clip]{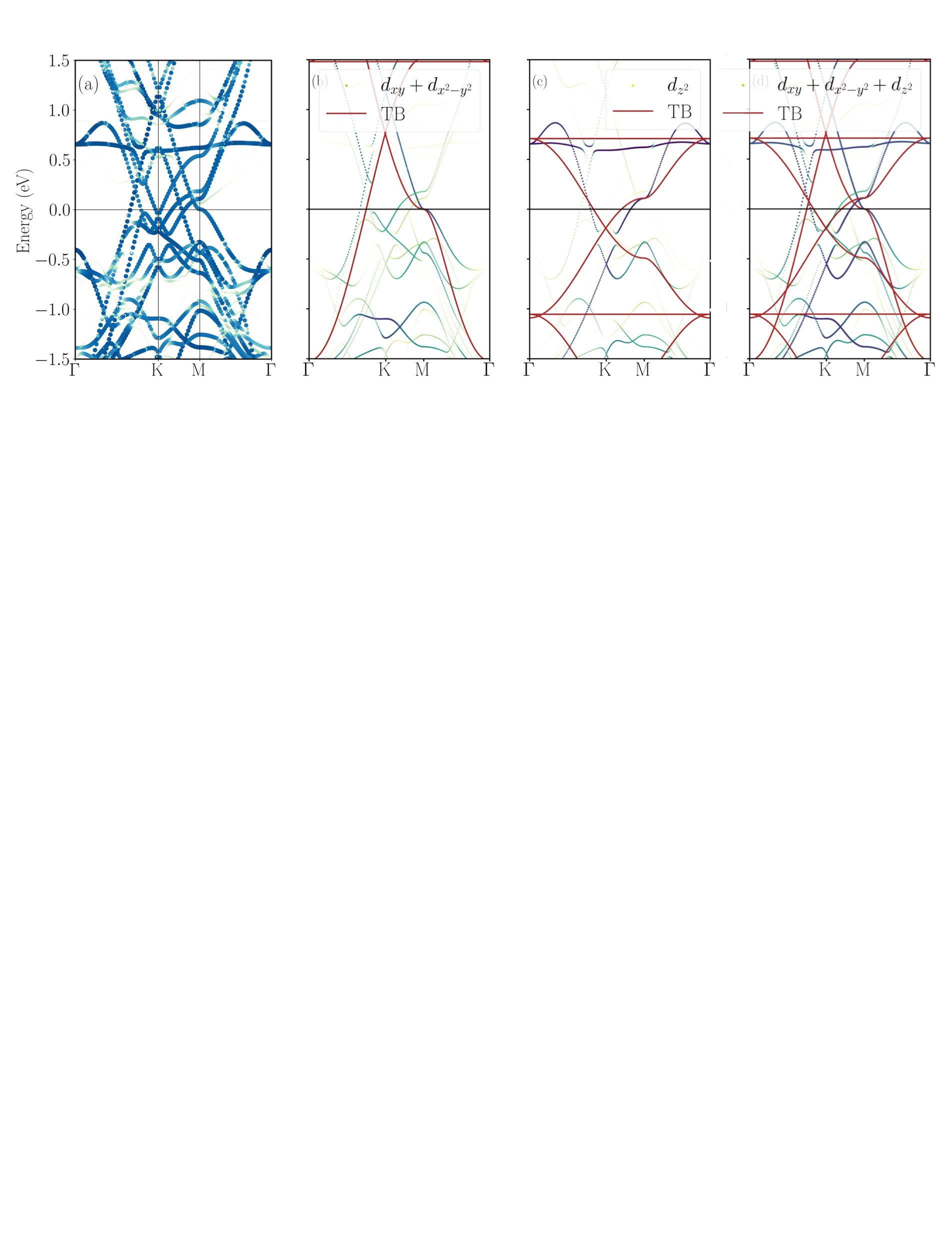}
\vspace{-1.0 cm}
\caption{{\textbf{Electronic structure of LuNb$_6$Sn$_6$.} (a) Unfolded band structure for LuNb$_6$Sn$_6$ in the CDW phase. (b) Comparison between Nb $d_{xy}$ + $d_{x^2-y^2}$ orbitally-projected DFT bands, in the pristine phase, and corresponding TB model. (c) Comparison between Nb $d_{z^2}$ orbitally-projected DFT bands, in the pristine phase, and corresponding TB model. (d) Comparison between Nb $d_{z^2}$ + $d_{xy}$ + $d_{x^2-y^2}$ orbitally-projected DFT bands, in the pristine phase, and the complete 12 bands TB model.}}
\label{f:CDW_Nb_kagome}
\end{figure*}

In this subection, we present first principles results for LuNb$_6$Sn$_6$, showing that the physics of ScV$_6$Sn$_6$ applies in this system as well. In Fig. \ref{f:kzonehalf_CDW}, we illustrate the $k_z$ dispersion within the CDW phase; one observes the gapping out of the dispersion along the $\Gamma$-A line and depletion of $k_z$ spectral weight due to the out-of-plane distortion of the Lu and Sn atoms. As in ScV$_6$Sn$_6$, this justifies the use of a two-dimensional kagome model within the CDW phase. Fitting the bands to a tight-binding model of the form Eq. \ref{Htb}, we find the parameters $t_x=0.76$, $t_z=-0.30$, $\varepsilon_x=1.50$ and $\varepsilon_z=-0.77$ (all units in eV); the comparison between the two-orbital model and the DFT bandstructure is shown in Fig. \ref{f:CDW_Nb_kagome}, illustrating good agreement. As in the case of ScV$_6$Sn$_6$, we find that the two vHS near the Fermi level are $p$-type in nature, so that our arguments regarding the propensity to excitonic order carry through in this case as well.

\newpage
\subsection{Results for non-CDW 166 materials (Ho,Tb)Nb$_6$Sn$_6$}

We contrast ScV$_6$Sn$_6$ and LuNb$_6$Sn$_6$ with other 166 kagome metals (Ho,Tb)Nb$_6$Sn$_6$, for which we find no CDW instability. Consequently, these systems have significant $k_z$ dispersion which is expected to suppress excitonic order, as $k_z^2$ enters the mean field excitonic Hamiltonian with the same form as a momentum-dependent Zeeman field in the Bogoliubov-de Gennes Hamiltonian for superconducting pairing. To describe the bandstructure of these materials we find it necessary to introduce $c$-axis hoppings $t^{(c)}_i$ with $i=x,z$; fitting tight-binding parameters for $A$Nb$_6$Sn$_6$, we find $\{\varepsilon_x,\varepsilon_z,t_x,t_z, t^{(c)}_x,t^{(c)}_z\}=\{1.68,-0.61,0.76,-0.30,0.75,0.44\}$ for $A$=Ho and $\{\varepsilon_x,\varepsilon_z,t_x,t_z, t^{(c)}_x,t^{(c)}_z\}=\{1.68,-0.61,0.76,-0.30,0.75,0.44\}$ for $A$=Tb. The significant $c$-axis dispersion in these systems distinguishes their physics from the CDW phase of Sc166 and Lu166.

\begin{figure*}[h!]
\centering
\includegraphics[width = 0.945\textwidth, trim={0.2cm 21cm 1.25cm 0cm},clip]{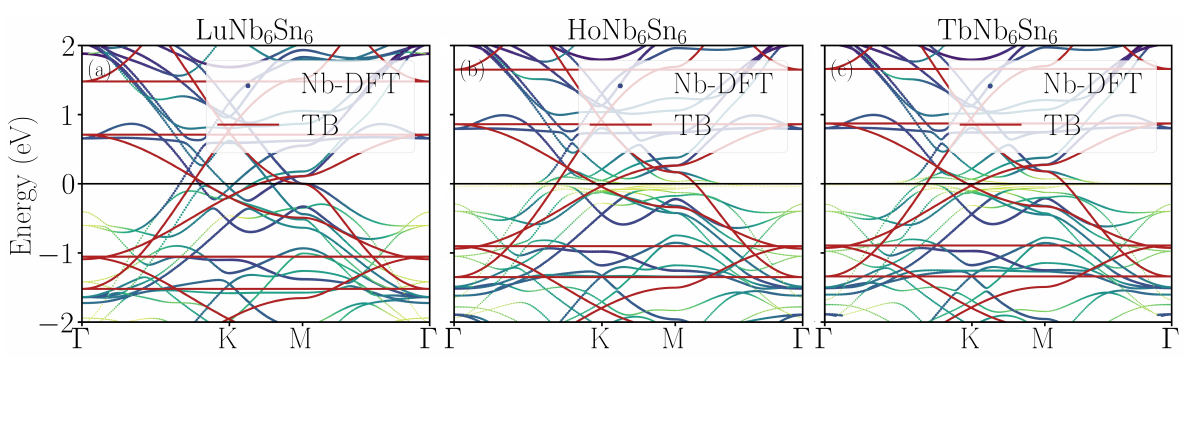}
\vspace{-1.0 cm}
\caption{{\textbf{Electronic structure of (Lu,Ho,Tb)Nb$_6$Sn$_6$.} Comparison between Nb $d_{z^2}$ orbitally-projected DFT bands in the pristine phase, and corresponding TB model, for LuNb$_6$Sn$_6$ (a), HoNb$_6$Sn$_6$ (b) and TbNb$_6$Sn$_6$.}}
\label{f:Nb_kagome}
\end{figure*}

\newpage
\section{Inter-orbital Hubbard model}
\begin{figure}[b]
    \centering
    \includegraphics[width=0.65\linewidth]{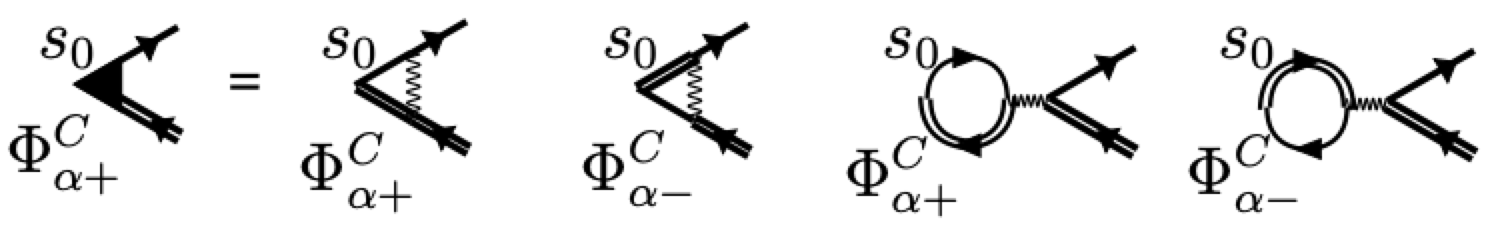}
    \caption{\textbf{Diagrammatic contributions to the excitonic pairing vertex.} Diagrammatic representation of the mean-field equations for spin-singlet excitonic order, with $\Phi^C_{\alpha +} = \langle c^\dag_{\alpha} d_{\alpha} \rangle$ and $\Phi^C_{\alpha +}=(\Phi^C_{\alpha -})^\dag$.}
    \label{fig:vertex}
\end{figure}
In this section we discuss the mean-field analysis for excitonic order from short-ranged interactions. The interactions which generate excitonic order are those which couple the opposite concavity vHS, which are purely composed of the $d_{z^2}$ and $d_{xy}/d_{x^2-y^2}$ orbitals respectively \footnote{The vHS due to the $d_{xz}$/$d_{yz}$ lie a distance $\sim$ 0.25 eV away from the Fermi level, and and so do not contribute significantly to the Fermi pockets near $M$.}. The interactions which drive excitonic order are therefore strictly those which couple the two distinct orbital types, which we index by $\alpha$ as in the main text. We write down an extended Hubbard model featuring only couplings between different orbitals, and between both electrons onsite and on neighbouring sites i.e. adjacent sublattices indexed by $\sigma$. We will use slightly different notation to the main text to better compare with prior work \cite{Scammell2023bib}; the $U'$ and $V'$ of the main text correspond to $U_{h_4}$ and $U_{l_2}$ below. The result is an interacting Hamiltonian,
\begin{gather}
H_\text{int} = \!\!\!\sum_{i,\braket{\sigma, \sigma'},\alpha\neq\alpha'} \bigg[ U_{h_1}(\psi^\dag_{i,\alpha,\sigma}\psi_{i,\alpha,\sigma'})(\psi^\dag_{i,\alpha',\sigma'}\psi_{i,\alpha',\sigma})   + U_{h_4}\delta_{\sigma,\sigma'}(\psi^\dag_{i,\alpha,\sigma}\psi_{i,\alpha,\sigma})(\psi^\dag_{i,\alpha',\sigma}\psi_{i,\alpha',\sigma})\\
   \notag      + U_{j_1}(\psi^\dag_{i,\alpha,\sigma}\psi_{i,\alpha',\sigma'})(\psi^\dag_{i,\alpha,\sigma'}\psi_{i,\alpha',\sigma}) + U_{j_2}  (\psi^\dag_{i,\alpha,\sigma}\psi_{i,\alpha',\sigma})(\psi^\dag_{i,\alpha,\sigma'}\psi_{i,\alpha',\sigma'}) 
     +  U_{j_4}\delta_{\sigma,\sigma'} (\psi^\dag_{i,\alpha,\sigma}\psi_{i,\alpha',\sigma})(\psi^\dag_{i,\alpha,\sigma}\psi_{i,\alpha',\sigma}) \\    
      \notag   +  U_{l_2}(\psi^\dag_{i,\alpha,\sigma}\psi_{i,\alpha',\sigma})(\psi^\dag_{i,\alpha',\sigma'}\psi_{i,\alpha,\sigma'}) +     U_{l_4} \delta_{\sigma,\sigma'}(\psi^\dag_{i,\alpha,\sigma}\psi_{i,\alpha',\sigma})(\psi^\dag_{i,\alpha',\sigma}\psi_{i,\alpha,\sigma})\bigg]
\end{gather}
Denoting the creation operators for electron and hole-like fermions in band basis as $c^\dag_{\bm k}$ and $d^\dag_{\bm k}$, the transformation to band basis is effected via the unitary transformation $u_{\bm k, +, \sigma}c^\dag_{\bm k} = \psi^\dag_{{\bm k},+,\sigma}$ and $u_{\bm k, -, \sigma}d^\dag_{\bm k} = \psi^\dag_{{\bm k},-,\sigma}$, as described in the main text. Fig. \ref{fig:vertex} represents the mean field equation for the excitonic order parameter $\Phi^C_{\bm{k}+} = \langle c^\dag_{\bm{k}}d_{\bm{k}}\rangle$, with $\Phi^C_{\alpha +}=(\Phi^C_{\alpha -})^\dag$. The corresponding vertices appearing in Fig. \ref{fig:vertex} acquire form factors from the transformation to band basis, which we denote as $F^{\mu\nu}_{\bm k_1,\bm k_2; \sigma_1,\sigma_2} = u^*_{\bm k_1, \mu, \sigma_1}u_{\bm k_2, \nu, \sigma_2}$, and using that $F^{++}_{\bm k_1,\bm k_2; \sigma_1,\sigma_2}=F^{--}_{\bm k_1,\bm k_2; \sigma_1,\sigma_2}\equiv F_{\bm k_1,\bm k_2; \sigma_1,\sigma_2}$ we will make use of $F_{\bm k_1, \bm k_1; \sigma_1, \sigma_1}F_{\bm k_2, \bm k_2; \sigma_2, \sigma_2}=|F_{\bm k_1, \bm k_2; \sigma_1, \sigma_2}|^2$. We may introduce shorthand notation for each vertex,
 \begin{align}
 \label{vertices}
 V^{++}_{\bm k_1,\bm k_2; \sigma_1,\sigma_2}&=\left[U_{h_4}\delta_{\sigma_1,\sigma_2}+U_{h_1}(1-\delta_{\sigma_1,\sigma_2})\right]|F_{\bm k_1, \bm k_2; \sigma_1, \sigma_2}|^2\\
 B^{++}_{\bm k_1,\bm k_2; \sigma_1,\sigma_2}&=\left[U_{l_4} {\ }\delta_{\sigma_1,\sigma_2}+U_{l_2}(1-\delta_{\sigma_1,\sigma_2})\right]|F_{\bm k_1, \bm k_2; \sigma_1, \sigma_2}|^2\\
 V^{+-}_{\bm k_1,\bm k_2; \sigma_1,\sigma_2}&=\left[U_{j_4}{\ }\delta_{\sigma_1,\sigma_2}+U_{j_1} (1-\delta_{\sigma_1,\sigma_2})\right] |F_{\bm k_1, \bm k_2; \sigma_1, \sigma_2}|^2\\
 B^{+-}_{\bm k_1,\bm k_2; \sigma_1,\sigma_2} &=\left[U_{j_4}{\ }\delta_{\sigma_1,\sigma_2}+U_{j_2} (1-\delta_{\sigma_1,\sigma_2})\right] |F_{\bm k_1, \bm k_2; \sigma_1, \sigma_2}|^2
\end{align}
with $V^{\gamma_1,\gamma_2}=V^{\gamma_2,\gamma_1}$ and $B^{\gamma_1,\gamma_2}=B^{\gamma_2,\gamma_1}$ for $\gamma=\pm$.
We distinguish the vertex contributions $V$, from the bubble diagrams $B$, and define the spinless particle-hole susceptibility $\Pi_{\bm k}$. Taken together, the self-consistent vertex equation can be cast as a simple eigenvalue problem,
\begin{align}
\label{verteq}
    \begin{pmatrix} \Phi^+_{\bm k_1}\\ \Phi^-_{\bm k_1} \end{pmatrix} 
    &=\sum_{\bm k_2, \sigma_1, \sigma_2}  \begin{pmatrix}  V^{++}_{\bm k_1,\bm k_2; \sigma_1,\sigma_2} -2B^{++}_{\bm k_1,\bm k_2; \sigma_1,\sigma_2} & V^{+-}_{\bm k_1,\bm k_2; \sigma_1,\sigma_2} -2B^{+-}_{\bm k_1,\bm k_2; \sigma_1,\sigma_2}\\ V^{-+}_{\bm k_1,\bm k_2; \sigma_1,\sigma_2} -2B^{-+}_{\bm k_1,\bm k_2; \sigma_1,\sigma_2} & V^{--}_{\bm k_1,\bm k_2; \sigma_1,\sigma_2} -2B^{--}_{\bm k_1,\bm k_2; \sigma_1,\sigma_2}  \end{pmatrix}  \Pi_{\bm k_2} \begin{pmatrix} \Phi^+_{\bm k_2}\\ \Phi^-_{\bm k_2} \end{pmatrix}.
\end{align}
It remains to diagonalize $\bar{\Phi}=\hat{M}\bar{\Phi}$; we can further simplify the matrix to bring it to a form more familiar in patch treatments of vHS \cite{Scammell2023bib, Maiti2013, Schulz1987, Furukawa1998, Classen2020, Nandkishore2012, Park2021, nab2024pomeranchuk},
\begin{gather}
(\hat{M})_{\bm k_1,\bm k_2;\sigma_1,\sigma_2}= \\
\begin{pmatrix}  (U_{h_4}-2U_{j_4})\delta_{\sigma_1,\sigma_2}+(U_{h_1}-2U_{l_2})(1-\delta_{\sigma_1,\sigma_2}) & -U_{j_4}\delta_{\sigma_1,\sigma_2}+(U_{j_1}-2U_{j_2})(1-\delta_{\sigma_1,\sigma_2})\\ -U_{j_4}\delta_{\sigma_1,\sigma_2}+(U_{j_1}-2U_{j_2})(1-\delta_{\sigma_1,\sigma_2}) &  (U_{h_4}-2U_{j_4})\delta_{\sigma_1,\sigma_2}+(U_{h_1}-2U_{l_2})(1-\delta_{\sigma_1,\sigma_2})  \end{pmatrix} |F_{\bm k_1, \bm k_2; \sigma_1, \sigma_2}|^2 \Pi_{\bm k_2}.\nonumber
\end{gather}

\begin{figure*}[t!]
    \centering
    \includegraphics[width=\textwidth]{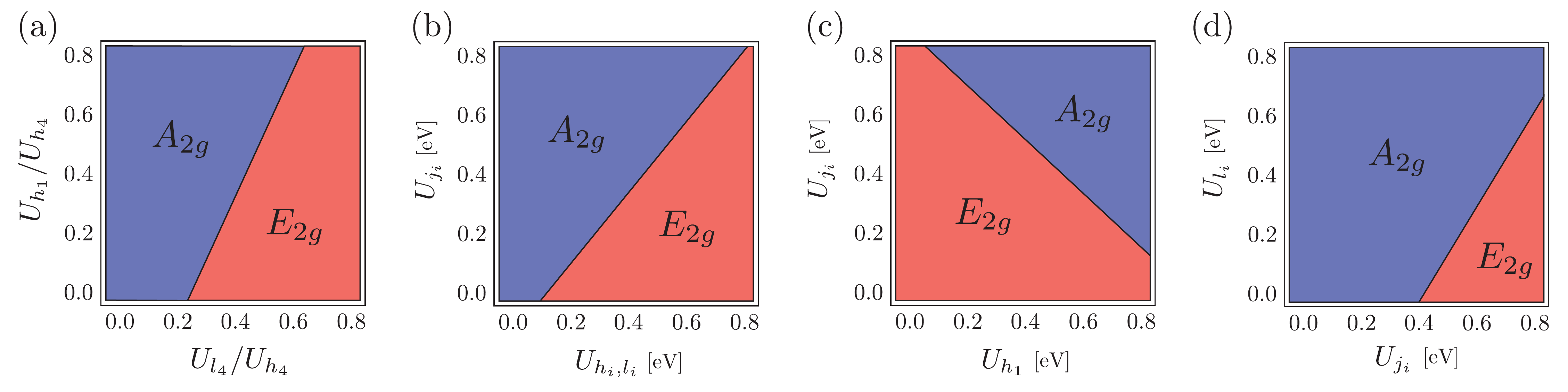}
    \vspace{-0.5cm}
    \caption{\textbf{Phase diagram as a function of Hubbard parameters.} Plot of the state with largest eigenvalue of the gap equation as a function of the Hubbard parameters; all units of energy are in eV. In the notation of the main text, $U_{l_2}=V'$ and $U_{h_4}=U'$. (a) $U_{h_1}$ versus $U_{l_4}$ at fixed $U_{h_4}=1$, with $U_{l_2}=U_{l_4}$ and $U_{j_1}=U_{j_2}=U_{j_4}=0.45$. (b) $U_{j_i}=U_{h_1}$ versus  $U_{l_i}$, for all $i$, at fixed $U_{h_4}=1$. (c) $U_{j_i}$ versus $U_{h_1}$, for all $i$, at fixed $U_{l_2}=U_{l_4}=0.5, U_{h_4}=1$. (d) $U_{l_i}$ versus $U_{j_i}$, for all $i$, at fixed $U_{h_4}=1, U_{h_1}=0.75$.  } 
    \label{phase_diags}
\end{figure*}

We explore the mean-field phase diagram by numerical diagonalization for various regions of parameter space. The key coupling which drives $E_{2g}$ excitonic order are the orbital exchange interactions, as explained in the main text; retaining only $U_{h_4}$ and $U_{l_2}$ produces the phase diagram shown in Fig. \ref{solutions} of the main text, in which small values of $U_{l_2}$ already result in $E_{2g}$ order. As illustrated by Fig. \ref{phase_diags}, various combinations of the other Hubbard parameters can promote the fully gapped $A_{2g}$ state. We leave ab initio estimates of these interaction parameters to future work.

In addition to the effects of strain, as well as the orbital exchange couplings which we have already discussed, perturbative renormalization group analyses have shown that fluctuation effects beyond mean field theory may also favour $E_{2g}$ order \cite{Scammell2023bib}, but such effects lie outside the scope of our present work.

\begin{figure*}[t!]
    \centering
    \includegraphics[width=0.9\textwidth]{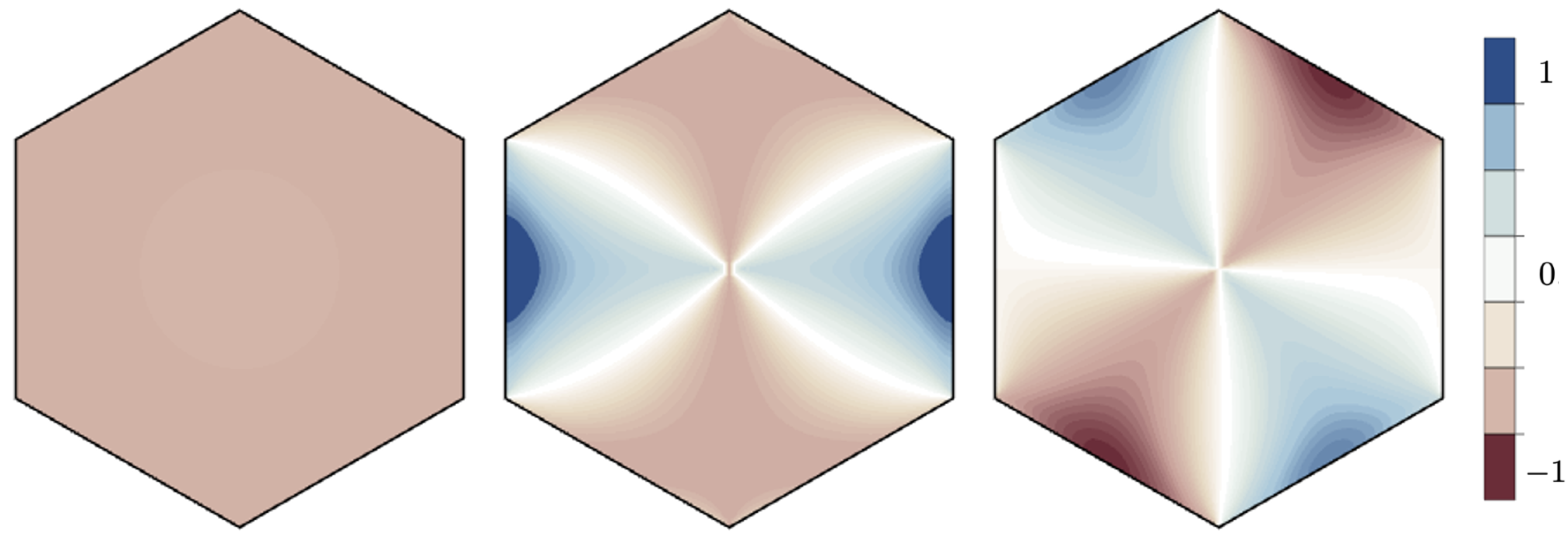}
    \caption{\textbf{Momentum space structure of the gap equation solutions.} The eigenstates of Eq. \eqref{verteq} for the three states with non-zero eigenvalue, for the case of $U_{h_4}\neq0$ and all other interactions zero (presented in arbitrary units). The corresponding (normalized) eigenvalues are $\{1, 0.44, 0.44\}$ and zero for all other states, but the momentum space structure of the eigenvectors themselves is essentially the same for different choices of parameters. The eigenstates are seen to transform as $\Gamma_\Phi=\{A_{1g}, E_{2g}^{(1)}, E_{2g}^{(2)}\}$, while the full order parameter transforms as $\Gamma=\Gamma_\Phi\otimes A_{2g} = \{A_{2g}, E_{2g}^{(1)}, E_{2g}^{(2)}\}$. } 
    \label{OPs}
\end{figure*}

The $E_{2g}$ can be understood in momentum space as corresponding to sign changes as one traverses through the $M$ points. Due to the $p$-type nature of each vHS, the $E_{2g}$ can therefore be understood as in real space as sign changes occurring between sublattices. To illustrate the momentum space structure of the solutions, we plot the eigenvectors of the mean-field equations throughout the full Brillouin zone in Fig. \ref{OPs} for $U_{h_4}=1$ and all other couplings zero. Note that the largest eigenvalue (corresponding to the lowest energy bound state) in this case is the $A_{1g}$ solution, but the $E_{2g}$ state is nonetheless present as a solution due to the momentum dependence of the form factors.

The fact that a subleading $E_{2g}$ order can arise even for the simple onsite $U'$ interaction derives from the effects of the wavefunction form factors, which can and do add spatially varying prefactors to otherwise spatially structureless couplings \cite{Li2020b, Scammell2021, Ingham2023}. We illustrate this point using analytical arguments. Consider exciton pairing in a patch about a given $M$ point, say $\bm M_1$ at which the wavefunctions are localized at A-sites. Using the approximate particle-hole symmetry, then
\begin{align}
\label{gapeqA}
    \notag \Phi^{\bm M_1}_{\bm k_1}&=U' \sum_{\bm k_2\in{\cal P}(\bm M_1)} {\cal U}^*_{\bm k_1, +, A}{\cal U}_{\bm k_2, +, A} {\cal U}^*_{\bm k_2, +, A}{\cal U}_{\bm k_1, +, A}  \Phi^{\bm M_1}_{\bm k_2}\\
    &=U' \sum_{\bm k_2\in{\cal P}(\bm M_1)} |{\cal U}_{\bm k_1, +, A}|^2 |{\cal U}_{\bm k_2, +, A}|^2  \Phi^{\bm M_1}_{\bm k_2}\\
    \label{gapeqA_approx}
   \Phi^{\bm M_1}_{\bm k_1} &\approx U' \sum_{\bm k_2\in{\cal P}(\bm M_1)}  \Phi^{\bm M_1}_{\bm k_2}.
\end{align}
We have used that $|{\cal U}_{\bm k, +, A}|^2\approx 1$ and $|{\cal U}_{\bm k, +, B/C}|^2\approx 0$ for $\bm k \approx \bm M_1$. The approximate gap equation \eqref{gapeqA_approx} clearly has solution $\Phi^{\bm M_1}_{\bm k_1} = \text{constant}$, and importantly, we see that it is even under all $C_{2}$ axes and mirrors. Accounting for all three patches, the eigenstates will decomposed into irreps of $C_{3z}$, giving just $A, E$. Taking into account the behaviour under the $C_2$ and mirrors, the $A, E$ irreps of the patch theory become $A_{1g}, E_{2g}$ of the continuous $D_{6h}$ theory.

We emphasize that, following the discussion in the main text, the full symmetry of the order parameter is not only determined by the $A$/$E$ form factor determined by solving the gap equation of our two-dimensional model, but the orbital content of the parent bands, which determine the symmetry behavior under out-of-plane symmetry operation $M_z$. The order parameter symmetry one expects in different kagome bilayer materials may therefore differ as a result of these considerations.

\end{document}